# Josephson vortex dynamic in superconducting $YBa_2Cu_3O_x$ ceramics

A.A.Sahakyan[*], S.K.Nikoghosyan, H.N.Yeritsyan, G.V.Grigoryan

Radiation Physics Laboratory, Yerevan Physics Institute, 2, Alikhanyan Bros. str.,

Yerevan 375036, Armenia

(Dated: June 22, 2004)

## Abstract

The Josephson vortices dynamic in superconducting $YBa_2Cu_3O_x$ ceramics was studied in frequency range 0.01-0.5 Hz by ac susceptibility measurements in non complete magnetic flux penetration regime.

We have found that with the increase of magnetic field frequency starting from 0.01 Hz the ac magnetic response shows complex frequency dependence: in one half period of a sinusoidal field appear two asymmetrical peaks of real $\Delta\chi'$ and imaginary $\chi''$ parts of the ac susceptibility. The analysis of experimental results has led to an idea that probably the volume (diameter) of Josephson vortices increases during its movement. The experimental results are discussed in terms of Josephson vortex - vortex interactions.

PACS numbers: 74.25.Ha; 74.60.Ge; 74.72.Bk

**1. Introduction**

In our previous work [1] we have reported on dynamic response of superconducting $YBa_2Cu_3O_x$ ceramics in magnetic field frequency range 0.01-90 Hz and amplitude of sinusoidal field up to 40 Oe. We have found, that in a regime of incomplete penetration of a sinusoidal magnetic field in intergranular medium of a superconducting sample some features are observed, from which it is possible point out several main results.

(i) The dynamic magnetic response in HTSC materials has nonlinear character and strong frequency dependence.

(ii) The Josephson vortices created by a static field and dynamic vortices created by sinusoidal ac field have different properties, in particular their density essentially differs.

We have assumed that vortex-vortex interaction plays an important role in dynamic properties of Josephson vortices and tried to explain the obtained experimental data basically from this point of view. In this paper we have attempted to observe experimentally the penetration dynamics of Josephson vortices in intergranular medium of superconducting $YBa_2Cu_3O_x$ sample. Obtained results show, that both components of magnetic ac susceptibility parameters have two important features. First, with the increase of magnetic field frequency starting from 0.01 Hz there are a constant and a variable components in a signal of real $\Delta\chi'$ and imaginary $\chi''$ parts of the ac susceptibility. Second, in one half-cycle of a sinusoidal field two asymmetrical peaks of susceptibility component $\Delta\chi'$ and $\chi''$ appear which increase with the increase of a frequencies or amplitudes of magnetic field.

In this paper we present experimental results on the Josephson vortices dynamics in intergranular medium of superconducting $YBa_2Cu_3O_x$ ceramics in non complete magnetic flux penetration regime in frequency range from 0.01 Hz to 0.5 Hz.

---

[*] Electronic address: asahak@mail.yerphi.am

## 2. Experimental

Cylindrically shaped (diameter-3 mm, length - 8 mm) ceramic $YBa_2Cu_3O_x$ samples were prepared by standard solid-state reaction technology. The real $\Delta\chi'$ and imaginary $\chi''$ parts of complex magnetic susceptibility of HTSC samples were measured by weak ac magnetic field susceptibility technique at frequency $f_m = 10 kHz$ in the presence of external low frequency ($f_{ext}$) sinusoidal magnetic field $H_{ext}(t) = H_0 \sin(2\pi f_{ext} t)$, where $t$ is the time. The measuring ac magnetic field $h_m(t) = h_0 \sin(2\pi f_m t)$ is considered weak if its contribution to the values of the susceptibility components $\Delta\chi'$ and $\chi''$ is negligible. For each sample and measuring regime there is a weak magnetic field $h_0$ amplitude range where both real and imaginary components of ac susceptibility are independent of $h_0$. On the other hand the decrease of $h_0$ is limited by the equipment sensitivity. Taking into account these two requirements the optimum of ac measuring field amplitude is at the high end of the weak magnetic field range. For our investigations $h_0 = 2 mOe$ was obtained.

Complex magnetic susceptibility was measured using a homemade ac inductance bridge working at frequency $f_m = 10 kHz$ with amplitude $h_0$ in range from 0.5 mOe to 10 Oe. The real $\Delta\chi'$ and imaginary $\chi''$ parts of the ac susceptibility are determined from the data of measuring coil ($L_m$) parameters deviation caused by a superconducting sample placed in this coil. The $L_m$ is connected in the arm of the ac inductance bridge (Fig.1). Another external larger coil $L_{ext}$ was used for creation of static $H_{dc}$ or sinusoidal $H_{ext}(t)$ magnetic field in frequency range from 0.01 Hz to 90 Hz and amplitude up to 40 Oe. The axes of both coils were directed parallel to the cylindrical sample axis. The phase angle adjustment of two Lock-in detectors for both real and imaginary signal components was carried out connecting and disconnecting a calibrating inductance $L_c$ and resistance $R_c$ in series to the measuring coil $L_m$. Our experimental installation has a time resolution of about 10 ms and allows the measurement of the average values of $\Delta\chi'$ and $\chi''$ and direct observation of the penetration dynamics of a low-frequency (0.01-0.5 Hz) magnetic field in volume of superconducting HTSC sample.

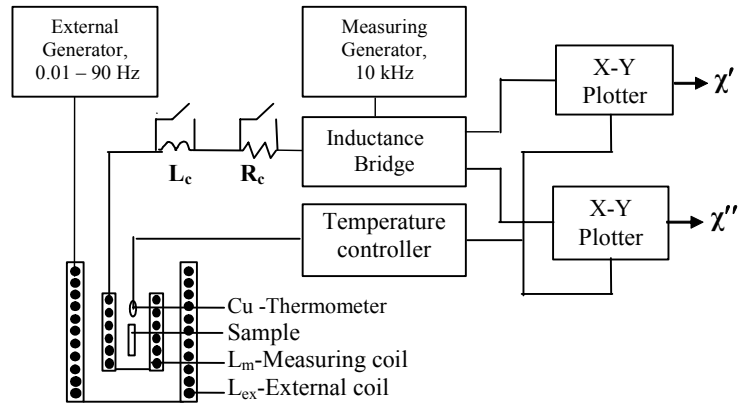

Fig.1.Schematic diagram of the ac magnetic susceptibility measurement installation: $L_c (80 \mu H)$ and $R_c (5\Omega)$ are calibrating inductance and resistance with button switch.

The real part of ac susceptibility has been taken as a deviation $\Delta\chi' = (\chi'_0 - \chi')/\chi'_0$, where $\chi'$ is the measured value of the real part of $\chi$ and $\chi'_0$ is the value of $\chi'$ at which a full screening state (Meissner state) of the sample occurs. The magnitude of $\Delta\chi'$ is proportional to the fraction of the sample magnetic flux penetration volume. For our measurements $\Delta\chi' \cong 0.28$ corresponds to the sample state at which flux front reaches the sample center [1].

Temperature of the sample was monitored with the help of a copper wire resistor with relative accuracy of about 0.1 K. The measuring coil $L_m$ was put in the liquid nitrogen.

### 3. Results and discussion

In Fig.2 illustrates the time dependences of the real $\Delta\chi'$ and imaginary $\chi''$ parts of the complex susceptibility of the ceramic sample $YBa_2Cu_3O_x$ measured for different ac field frequency and at $H_0 = 40 Oe$ amplitude of external sinusoidal field $H_{ext}(t)$. Apparently, at frequency 0.01 Hz both $\Delta\chi'$ and $\chi''$ components are directly proportional to absolute value of sinusoidal field $|H_{ext}(t)|$, i.e. the linear dynamic response is observed. However, with the increase of frequency the $\Delta\chi'$ and $\chi''$ shows two important features.

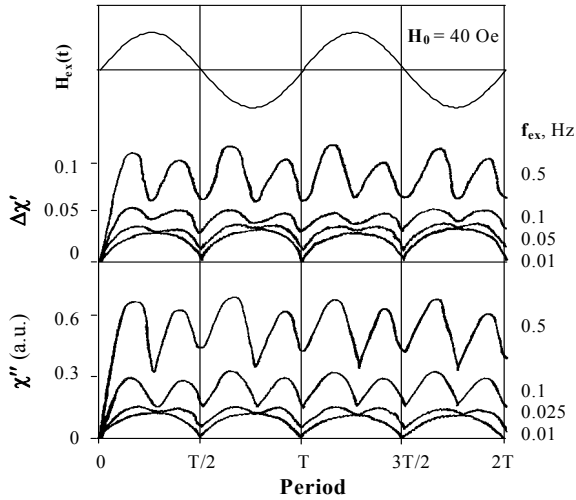

Fig.2. Evolution of the time dependences of the real $\Delta\chi'$ and imaginary $\chi''$ parts of the ceramic $YBa_2Cu_3O_x$ ac susceptibility for amplitude of $H_0 = 40 Oe$ at different external magnetic field frequency. The measuring temperature was T=80 K.

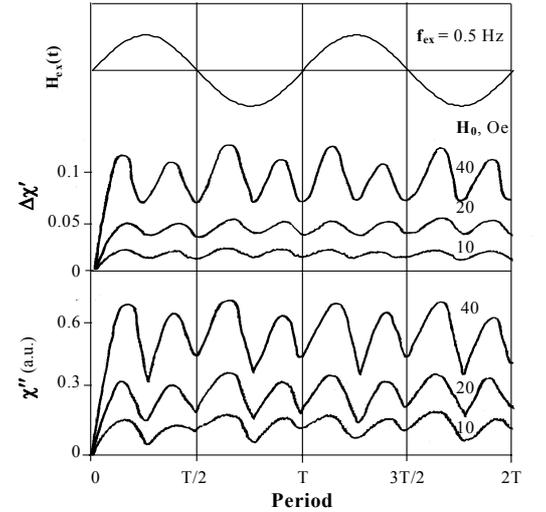

Fig.3. Evolution of the time dependences of the real $\Delta\chi'$ and imaginary $\chi''$ parts of the ceramic $YBa_2Cu_3O_x$ ac susceptibility at $f_{ext} = 0.5 Hz$ for different external magnetic amplitude. The measuring temperature was T=80 K.

First, with the increase of frequency asymmetric peaks (similar to humps) appear on the linear response of the sinusoidal form around the points $t = kT/8$ $(k = 2n+1, n \in N)$ appears. Here the $T$ is the period of external sinusoidal magnetic field. For convenience we shall call this "hump" effect. From these data it is possible suggest, that values of $\Delta\chi'$ and $\chi''$ are proportional to rate of a magnetic field change, i.e. $dH_{ext}(t)/dt$.

Second, apart from with a variable component of a signal of $\Delta\chi'$ and $\chi''$ a constant part of these also appears, with the relative share of the latter increasing with frequency $f_{ext}$.

In Fig.3 illustrates the time dependences of the real $\Delta\chi'$ and imaginary $\chi''$ parts ac susceptibility of sample $YBa_2Cu_3O_x$ measured for several ac external sinusoidal field amplitude $H_0$ at frequency $f_{ext} = 0.5 Hz$. In this case also the above described effects are observed: the "hump" and constant components of $\Delta\chi'$ and $\chi''$ appear. With increase of the magnetic field amplitude $H_0$ both effects increase. From here one can deduce, that values of components $\Delta\chi'$ and $\chi''$ are proportional also amplitudes of a magnetic field $H_o$.

Thus, one can assume that a contribution of above described effects into the values of $\Delta\chi'$ and $\chi''$ are proportional simultaneously to both parameters of a magnetic field: the rate of the magnetic field variation $dH_{ext}(t)/dt$ and the amplitude $H_o$. Hence the observed increases of ac susceptibility components are proportional to $|H_{ext}(t) \cdot dH_{ext}(t)/dt| \propto |dH_{ext}^2/dt|$ i.e. are proportional to time derivative of the magnetic field energy.

There should be also a contribution independent of the field changing rate, since for very slowly changing magnetic fields the values of $\Delta\chi'$ and $\chi''$ differ from zero. Hence, the dependences of $\Delta\chi'$ and/or $\chi''$ components can be presented as follows.

$$\chi'' \propto |H_{ext}(t)| + \alpha''|dH_{ext}^2(t)/dt| \quad (1)$$

where $\alpha''$ -is a calibration parameter. If we take $H_{ext}(t) = H_0 \sin(\omega_{ext}t)$, where $\omega_{ext} = 2\pi f_{ext}$, then

$$\chi'' \propto |H_0 \sin(\omega_{ext}t)| + \alpha'' H_0 \omega_{ext} |H_0 \sin(2\omega_{ext}t)| \quad (2)$$

The comparison of the experimental curve for time dependence of hysteresis losses $\chi''$ with the curve calculated using (2) (see Fig.4) favors such assumption. We see that apparent similarity of these curves. Significant discrepancies between them are observed at points $t = nT/2$ where the magnetic field is equal to zero. This is possibly connected with not taking into account Josephson vortices viscous drag in (2), which doesn't allow the vortices to leave the sample volume completely when the frequency of a field increases. In the following half-cycle of a sinusoidal field when the field direction is antiparallel to vortices still remaining in the volume of a sample, the latter considerably easier will penetrate into volume of a sample because of Josephson vortex-vortex attraction.

One can also see some difference in "hump" effect for components $\Delta\chi'$ and $\chi''$ on fig.2 and fig.3. It is possibly due to the fact, that $\chi''(t)$ describes energy losses on the movement of vortices, while $\Delta\chi'$ describes the volume of a sample containing these vortices. Thus the minima of $\chi''(t)$ at points $t = (n+1)T/2$ and those at $t = kT/4$ should differ since the magnetic field in the first case passes through a zero and changes the direction. I.e., in this case the number of moving vortices and their path lengths are significantly greater than those around the point $t = kT/4$, when movement of vortices is caused only by appearance of the "hump" effect. Hence, losses of energy at points $t = (n+1)T/2$ will be higher.

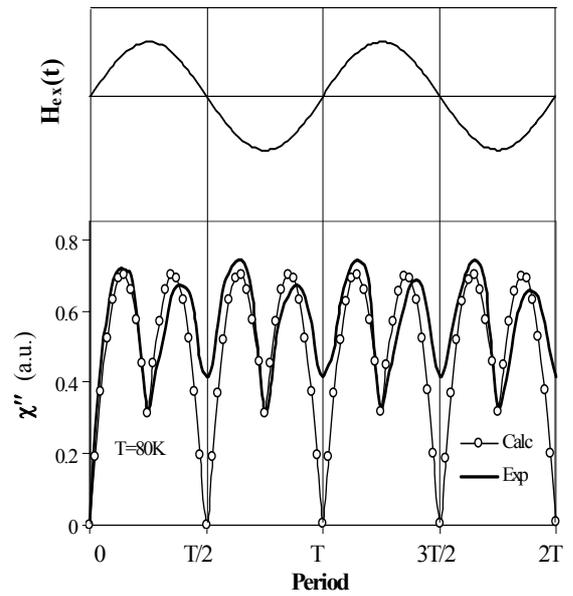

Fig.4. Time dependence of hysteresis losses $\chi''$ calculated by (2) and experimental curves measured at frequency $f_{ext} = 0.5 Hz$ and amplitude $H_0 = 4 Oe$ ($\alpha'' = 1.2 \cdot 10^{-2}$).

It is possible to explain the asymmetry of peaks of experimental curves in the "hump" effect from the point of view Josephson vortex-vortex interactions, here the interaction between frozen and sample penetrating vortices. In some cases they are parallel, and, hence, they repel, and thus the vortices penetration into sample is relative small, and in the other cases they are attracted and penetrate deeper.

We investigated also the evolution of the temperature dependence of the granule hysteresis losses peak $\chi''$ for some frequencies of a magnetic field for the amplitude $H_0 = 5 Oe$.

The temperature dependence of $\chi''$ is shown in Fig.5. Note that with increase of the magnetic field frequency a significant increase in the height of peak is observed, i.e. hysteresis losses on Abrikosov vortices increase. These data allow to assume, that possibly the behaviours of Abrikosov vortices are similar those of Josephson vortices. Note, that strong frequency dependences of magnetic susceptibility $\Delta\chi'$ and $\chi''$ was found in superconducting YBaCuO monocrystal [2]. Also for ac susceptibility in a near of the transition region of YBaCuO sintered samples quite complex dependence on the frequency and amplitude of the ac magnetic field was observed [3].

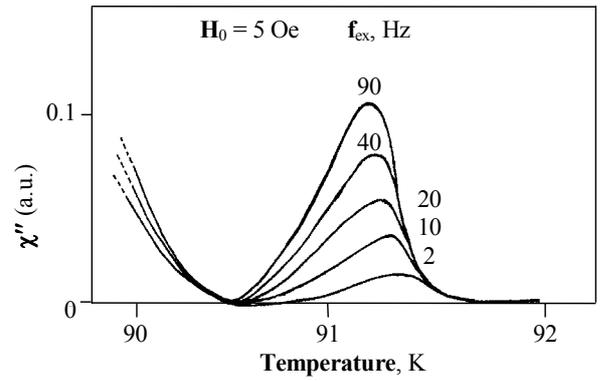

Fig.5. Temperature dependence of intragranular hysteresis losses $\chi''$ of ceramic $YBa_2Cu_3O_x$ for several magnetic field frequencies.

Summing up we can say, that all data on frequency and field investigations of superconducting YBaCuO ceramic samples can be explained phenomenological if one assumes, that the volume of moving Josephson vortices increases. Assuming, that vortex has the cylindrical form, and its axis is parallel to a magnetic field, the dependence of the vortex diameter ($d$) on the frequency of magnetic field can be described in the following way: $d(f_{ext}) \propto (\Delta\chi'(f_{ext}))^{1/2}$. In our previous work [1] we have found that relative increase of $\Delta\chi'$ at the sample temperature of an 80 K was about 60, in the case, when the vortices reach the center of a sample. Hence, presumably, the diameter of a vortex in this case increases more than 7 times.

Note that "hump" effect was observed in superconducting ceramic YBaCuO samples of any form: the cubic form, a plate and even a shapeless sample, and also for bismuth based ceramic HTCS sample ($Bi_{1.7}Pb_{0.2}Sb_{0.1}Sr_2Ca_2Cu_3O_x$; $T_c = 108$ K).

## 4. Conclusions

The investigations of Josephson vortices dynamics in superconducting $YBa_2Cu_3O_x$ ceramics in the frequency range 0.01-0.5 Hz by ac susceptibility measurements in non complete magnetic flux penetration regime has shown, that susceptibility parameters $\chi'$ and $\chi''$ have complex frequency dependence.

We have found that the real and imaginary parts of the ac magnetic susceptibility show two important features depending on frequency and amplitude of a magnetic field: (i) with increase of frequency (starting from 0.01 Hz and above) constant and variable components for magnetic susceptibility components $\Delta\chi'$ and $\chi''$ appear and (ii) in one half-cycle of a sinusoidal field two asymmetrical peaks of susceptibility component $\Delta\chi'$ and $\chi''$ appear.

The analysis of experimental results has led to idea, that, probably, the volume (diameter) of a moving Josephson vortex increases.

Temperature dependence of intragranular losses in frequency range from 2 Hz to 90 Hz shows that probably the behaviors of Abrikosov vortices are similar to those of Josephson vortices.